# Precise Schedulability Analysis for unfeasible to notify separately for comprehensive - EDF Scheduling of interrupted Hard Real-Time Tasks on the similar Multiprocessors

Jagbeer Singh
Dept. of Computer Science and Engineering, Gandhi Institute of Engineering and Technology Gunupur, Rayagada (Orissa), India-765022,
E-mail:- willybokadia@gmail.com , Mob No. +919439286506, +917873887979

## *Abstract*

In Real-time system, utilization based schedulability test is a common approach to determine whether or not tasks can be admitted without violating deadline requirements. The exact problem has previously been proven intractable even upon single processors; sufficient conditions are presented here for determining whether a given periodic task system will meet all deadlines if scheduled non-preemptively upon a multiprocessor platform using the earliest-deadline first scheduling algorithm. Many real-time scheduling algorithms have been developed recently to reduce affinity in the portable devices that use processors. Extensive power aware scheduling techniques have been published for energy reduction, but most of them have been focused solely on reducing the processor affinity. The non-preemptive scheduling of periodic task systems upon processing platforms comprised of several same processors is considered.

## 1 Introduction

Based on the functional criticality of jobs, usefulness of late results and deterministic or probabilistic nature of the constraints, the real time systems are classified as, *Hard real-time system* in which consequences of not executing a task before its dead line catastrophic or fatal, *Soft real-time system* in which the utility of results produced by a task decreases over time after deadline expires and *Firm* or *Weakly hard real-time system* in which the result produced by a task ceases to be useful as soon as the deadline expires but the consequences of not meeting the deadline are not very severe [1]. Typical illustrating examples of systems with weakly-hard real time requirements are multimedia systems in which it is not necessary to meet all the task deadlines as long as the deadline violations are adequately spaced. Computations occurring in a real-time system that have timing constraints are called real-time tasks. A real-time application usually consists of a set of cooperating tasks activated at regular intervals and/or on particular events. Tasks in real-time system are of two types, periodic tasks and aperiodic tasks [1]. Periodic tasks are time driven and recur at regular intervals called the period. Aperiodic tasks are event driven and activated only when certain events occur. The necessary condition is that real-time tasks must be completed before their deadlines for a system to be successful.

The problem of scheduling such tasks upon a single processor (CPU) so that all the deadlines are met has been widely studied in the literature and is now well understood. The most important point in this direction being that an optimal online scheduler, commonly known as *Earliest Deadline First* (EDF), has been derived. Earliest-deadline first is a priority-based scheduler which assigns priorities to jobs so that the shorter the absolute deadline of a job the higher its priority. This scheduler is optimal with the interpretation that if a periodic constrained-deadline task system can be successfully scheduled with another scheduler upon a single CPU, then it can also be successfully scheduled using earliest-deadline first. However, a very large number of applications nowadays turns out to be executed upon more than one CPU for practical and economic reasons due to the advent of multicore technologies. For such applications, even though earliest-deadline first is no longer optimal [8], much recent work gave rise to multiple investigations and thus many alternative algorithms based on this scheduling policy have been developed due to its optimality upon uniprocessor platforms [13]. Most results have been derived under either comprehensive or partitioned scheduling techniques. Over the years, the preemptive periodic constrained-deadline task model [9] has proven remarkably useful for the modeling of recurring processes that occur in hard real-time computer application systems, where the failure to satisfy any constraint may have disastrous consequences.

In comprehensive scheduling [4], all the tasks are stored in a single priority-ordered queue and the comprehensive scheduler selects for execution the highest priority tasks from this queue. In this framework, tasks are allowed to migrate at runtime from one CPU to another in order to complete their executions [6, 3]. Regarding this kind of schedulers, an important issue consists in deriving an precise schedulability test by exploiting on the one hand the predictability property of the scheduler and by providing on the other hand a feasibility interval so that if it is possible to find a valid schedule for all jobs contained in this interval, then the whole system will be stamped feasible. In partitioned scheduling [2], all the tasks are first assigned statically to the CPUs, and then each CPU uses independently its local scheduler at runtime. Despite these two scheduling techniques are incomparable [1] in the sense that there are systems which are schedulable with partitioning and not by comprehensive and conversely, and despite the high number of interesting results that have already been derived up to now, many open questions still remain to be answered, especially when comprehensive schedulers are considered.

***Associated work*** In recent years, as most comprehensive schedulers are predictible, extensive efforts have been performed towards addressing the problem of determining a feasibility interval for the comprehensive scheduling of periodic constrained deadline tasks upon multiprocessor platforms. That is, to derive an interval of time so that if it is possible to find a valid schedule for all jobs contained in this interval, then the whole system is feasible. Up to now, sound results have been obtained only in the particular case where tasks are scheduled by using a *Fixed-Task-Priority* (FTP) scheduler [6, 14]. Being an FTP scheduler one where all the jobs belonging to a task are assigned the same priority as the priority assigned to the task beforehand (i.e., at design time). We are not currently aware of any existing result concerning the feasibility interval for Fixed-Job-Priority (FJP) schedulers in the literature, except the one proposed by Leung in [12]. However, we show that this result is actually wrong. An FJP scheduler is one where two jobs belonging to the same task may be assigned different priorities.

***Purposed Research:*** To the best of my knowledge, this will be the first valuable feasibility interval for Fixed-Job-Priority schedulers since the one proposed by Leung in [12] is flawed. Based on this feasibility interval and considering the predictability property of this scheduler, our main contribution is therefore a precise schedulability test for the comprehensive-Earliest deadline first scheduling of periodic hard real-time tasks upon same multiprocessor platforms. In this paper, we derive a feasibility interval for a Fixed-Job-Priority scheduler, namely comprehensive-Earliest deadline first.

## 2 System Model

In this section, we briefly discuss the processor preemptive and non-preemptive scheduling and task models that we have used in our work. Throughout this paper, all timing characteristics in our model are assumed to be non-negative integers, i.e., they are multiples of some elementary time intervals (for example the CPU tick, the smallest indivisible CPU time unit for individual processor).

### 2.1 Task specifications

We consider the preemptive scheduling of a hard real-time system $\tau = \{\tau_1, \tau_2, ....... \tau_n\}$ composed of *n* tasks upon *m* same CPUs according to the following interpretations.

***Preemptive and non-preemptive scheduling:*** *- CPU scheduling decision may take place under the following four circumstances:-*
  *(i) When a process switch from the running state to the waiting state (e.g, I/O request,*

*or invocation of wait for the termination of one of the child processes)*

(ii) *When a process switches from the running state to the ready state (e.g, when an interrupt occurs)*

(iii) *When a process switches from the waiting state to the ready state(e.g, completion of I/O)*

(iv) *When a process terminates*

*So, when scheduling takes place only under circumstances (i) and (iv), we say the scheduling scheme is **non-preemptive**; otherwise, the scheduling is **preemptive**. Or we can define in short preemptive scheduling is," an executing task may be interrupted at any instant in time and have its execution resumed later".*

**Same CPUs:** *all the CPUs have the same computing capacities.*

We assume that all the tasks are independent, i.e., there is no communication, no precedence constraint and no shared resource (except for the CPUs) between tasks. Also, we assume that any job $\tau_{i,j}$ cannot be executed in parallel, i.e., no job can execute upon more than one CPU at any instant in time.

Each task $\tau_i$ is a periodic constrained-deadline task characterized by four parameters ($O_i$, $C_i$, $D_i$, $T_i$) where $O_i$ is the first release time (offset), $C_i$ is the Worst Case Execution Time, $D_i \leq T_i$ is the relative deadline and $T_i$ is the period, i.e., the Precise inter-arrival time between two consecutive releases of task $\tau_i$. These parameters are given with the interpretation that task $\tau_i$ generates an infinite number of successive jobs $\tau_{i,j}$ from time instant $O_i$, with execution requirement of at most $C_i$ each, the $j^{th}$ job which is released at time $O_{i,j} = O_i + (j-1)$. $T_i$ must complete within ($O_{i,j}$, $d_{i,j}$) where $d_{i,j} = O_{i,j} + D_i$, the absolute deadline of job $\tau_{i,j}$.

Job $\tau_{i,j}$ is said to be active at time *t* if and only if $O_{i,j} < t$ and $\tau_{i,j}$ is not completed yet. More precisely, an active job is said to be running at time *t* if it has been allocated to a CPU and is being executed. Otherwise, the active job is said to be ready and is in the ready queue of the operating system. We assume without any loss of generality that $O_i \geq 0$, $\forall_i \in (1,2,\ldots n)$ and we denote by $O_{max}$ the maximal value among all task offsets, i.e., $O_{max} = max(O_1, O_2, \ldots O_n)$. We denote by *P* the *hyperperiod* of the system, i.e., the least common multiple (lcm) of all tasks periods: $P = lcm(T_1, T_2, \ldots T_n)$. Also, we denote by $C_\tau$ the sum of the Worst Case Execution Time of all tasks in $\tau$:

$$C\tau = \sum_{i=1}^{n} Ci$$

**2.2 Scheduler specifications**

We assume in this research that the preemptions and migrations of all tasks and jobs in the system are allowed at no cost or penalty. Based on this feasibility interval and considering the predictability property of this scheduler, our main contribution is therefore a precise schedulability test for the comprehensive-Earliest deadline first scheduling of periodic hard real-time tasks upon same multiprocessor platforms. In this paper, we derive a feasibility interval for a Fixed-Job-Priority scheduler, namely comprehensive- Earliest deadline first. We consider that tasks are scheduled by using the Fixed-Job-Priority (FJP) scheduler comprehensive-earliest-deadline first . That is, the following two properties are always satisfied: (i) the shorter the absolute deadline of a job the higher its priority and (ii) a job may begin execution on any CPU and a preempted job may resume execution on the same CPU as, or a different CPU from, the one it had been executing on prior to preemption.

## 3 Definitions and Properties

First, we formalize the notions of synchronous and asynchronous systems, schedule and valid schedule, and configuration. In this section we provide definitions and properties that will help us establishing our precise schedulability test.

***Def.1 Valid schedule:*** *A schedule σ of a task system τ = (τ$_1$, τ$_2$,,,,,, τ$_n$) is said to be valid if and only if no task in τ ever misses a deadline when tasks are released at their specified released times.*

***Def.2 Deterministic schedulers***: *A scheduler is said to be deterministic if and only if it generates a unique schedule for any given set of jobs.*

***Def.3 Work-conserving schedulers***: *A scheduler is said to be work-conserving if and only if it never idles a CPU while there is at least one active ready task.*

***Def.4 Synchronous systems***: *A task system τ = (τ$_1$, τ$_2$,,,,,, τ$_n$) is said to be synchronous if each task in τ has its first job released at the same time-instant c, i.e., $O_i = c$ for all $1 \leq i \leq n$. Otherwise, τ is said to be asynchronous.*

***Def.5 Schedule σ (t):*** *For any task system τ = (τ$_1$, τ$_2$,,,,,, τ$_n$) and any set of m same CPUs {π$_1$,π$_2$,.... π$_m$}, the schedule σ (t)of system τ at time-instant t is defined as σ : N → f{1,2,..... n}$^m$ where σ(t)= (σ$_1$(t),σ$_2$(t),...... σ$_m$(t)) with σ$_j$(t) =8>0; if there is no task scheduled on π$_j$ at time-instant t i; if task τ$_i$ is scheduled on π$_j$ at time-instant t:*

***Def.6 A-feasibility***: *A periodic constrained-deadline task system τ is said to be A-schedulable upon a set of m same CPUs if all the tasks in τ meet all their deadlines when scheduled using scheduler A, i.e., scheduler A produces a valid schedule.*

***Def.7 Predictability:*** *A scheduler A is said to be predictable if the A-feasibility of a set of tasks implies the A-feasibility of another set of tasks with same release times and deadlines, but smaller execution requirements. Before we present the main result of this paper, we need to introduce the following notations and results taken from [10] and [6].*

***Lemma 1** (Ha and Liu [10]). Any work-conserving and FJP scheduler is predictable upon same multiprocessor platforms.*

***Appreciation to Lemma 1***, we are guaranteed that the comprehensive- earliest-deadline first scheduler is predictable. Indeed, comprehensive- earliest-deadline first is a work-conserving and FJP scheduler. Thereby, given a periodic constrained-deadline task system τ , we can always assume an instance of τ in which all jobs execute for their whole Worst Case Execution Time. This leads us to consider hereafter a system having known jobs release times, deadlines and execution times. If S worst is the valid schedule obtained with these parameters by using the comprehensive- earliest-deadline first scheduler, then we are guaranteed to successfully schedule every other possible instance of τ in which jobs can execute for less than their Worst Case Execution Time by using the same comprehensive- earliest-deadline first scheduler.

*Lemma 2* (Cucu and Goossens [6]). Let S be the schedule of a periodic constrained-deadline task system τ constructed by using the comprehensive- earliest-deadline first scheduler. If the deadlines of all task computations are met, then S is periodic from some point with a period equal to P.

*Lemma 3* (Inspired from Cucu and Goossens [14]). Let S be the schedule of a periodic constrained-deadline task system □ constructed by using the comprehensive- earliest-deadline first scheduler. Then, for each task $τ_i$ and for each time instant $t_1 ≥ O_i$, we have $e_{i,t_1} ≥ e_{i,t_2}$, where $t_2 = t_1 + P$.

## 4 Precise Schedulability Test

In this section we provide a precise schedulability test for the comprehensive-earliest-deadline first scheduling of periodic hard real-time tasks upon same multiprocessor platforms. It is worth noticing that we assume in this section that each job of the same task (say $τ_i$) has an execution requirement which is exactly $C_i$ time units thanks to the predictability property of this scheduler. Based on the later result, the intuitive idea behind our approach is to construct a schedule by using an implementation of comprehensive- Earliest Deadline First which follows hypothesis described in Section 3, then check to see if the deadlines of all task computations are met. However, for this method to work we need to establish an "a priori" time interval within which we need to construct the schedule. If the task system τ is synchronous, then such a time interval is known: (0, *P*) where $P = lcm(T_1, T_2, ..... T_n)$ see [6] for details. Unfortunately, if the task system τ is asynchronous, such a time interval is unknown, in the following we will fill the gap.

As the task system τ is composed of periodic tasks, the idea thereby consists in simulating the system until the schedule becomes periodic, i.e., the steady phase representing the general timely behavior of the system from a certain time instant is reached. This steady phase is reached when two configurations separated by P time units are same.

*Study1*. By extending the results obtained in the uniprocessor framework to the multiprocessor platforms, Leung claimed in [12] that an Precise feasibility condition for comprehensive-earliest-deadline first consists in checking if (i) every deadline is met until time $O_{max} + 2P$ and (ii) the configurations at instants $O_{max} + P$ and $O_{max} + 2P$ are same. Anyway, this is flaw, since there are schedulable task systems that reach their steady phase later than $O_{max} + 2P$, as shown by example 1 taken from [5].

*Example 1* (Braun and Cucu [5]). Consider the following periodic task system: $\tau_1 = (O_1 = 0, C_1 = 2, D_2 = T2 = 3)$, $\tau_2 = (O_2 = 4, C_2 = 3, D_2 = T_2 = 4)$, $\tau_3 = (O_3 = 1, C_3 = 3, D_3 = T_3 = 6)$ to be scheduled with comprehensive- earliest-deadline first upon $m = 2$ CPUs.

By building the schedule (see Figure 1)[14], it is possible to see that at time $O_{max}+P = 4+12 = 16$ and $O_{max}+2P = 4 + 2*12 = 28$ the steady phase has not yet been reached (at times 114 and 29 there are two different configurations). However, the steady phase is reached after a further hyperperiod. Since no deadline is missed, the task system is schedulable with comprehensive- Earliest Deadline First.

The flaws in the results proposed by Leung in [12] come from many sources. The paper was actually centered on the Least Laxity First (LLF) scheduler defined as follows.

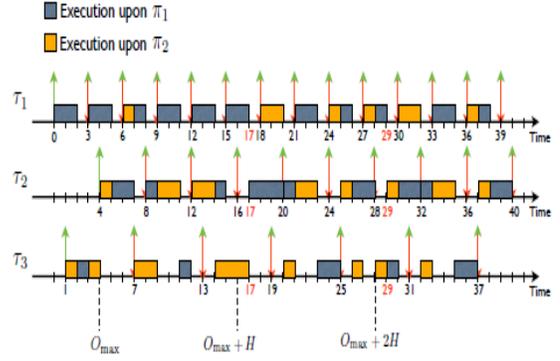

Figure 1: $[0, O_{max} + 2P]$ is not a feasibility interval for the global-EDF scheduler.

*Def.8 LLF scheduler*: The LLF scheduler always executes the jobs with least laxity; being the laxity of a job its absolute deadline minus the sum of its remaining processing time and the current time.

*Lemma 4* (Lemma 2, pages 216–2114 of [12]). Let S be the schedule of a task system $\tau$ constructed by using the comprehensive-Earliest Deadline First scheduler upon $m \geq 1$ CPUs. If $\tau$ is schedulable by using the comprehensive- Earliest Deadline First scheduler on m CPUs, then $CS(\tau, t_1) = CS(\tau, t_2)$ where $t_1 = O_{max} + P$ and $t_2 = t_1 + P$.

the *m* CPUs are always busy in the interval $[t_1, t_2]$ is incorrect; this is a uniprocessor argument not valid in a multiprocessor context. Indeed, considering example 1, it is not difficult to see in Figure (1) that $t_1 = 16$, $t_2 = 28$ and $C_S(\tau,16) \neq C_S(\tau,28)$. However, in the time-slots [114,18) and [23,24), only one CPU (here, CPU $\pi_1$) out of two is actually busy by the execution of the jobs.

*Observation:* Although we address the comprehensive- Earliest Deadline First scheduling problem of periodic constrained-deadline task systems, example 2 give evidence of possibly late occurrence of the steady phase in the valid schedule $S$ of a task system $\tau$. Indeed, it shows that the steady phase can be reached after a time-instant as large as $O_{max} + 42*P$.

*Example 2.* Consider the following periodic task system: $\tau_1 = (O_1 = 225, C_1 = 90, D_2 = T_2 = 161)$, $\tau_2 = (O_2 = 115, C_2 = 40, D_2 = T_2 = 161)$, $\tau_3 = (O_3 = 0, C_3 = 142, D_3 = T_3 = 161)$, $\tau_4 = (O_4 = 129, C_4 = 120, D_4 = T_4 = 161)$ to be scheduled with comprehensive- Earliest Deadline First upon $m = 2$ CPUs.

By building the schedule using an open source simulation tool such as STORM1 (we implemented a deterministic and request-dependent Earliest Deadline First tie-breaker), it is possible to see that at time-instants $O_{max} + 42P = 69814$ and $O_{max} + 43P = 14148$ the steady phase has not been reached yet (there are two different configurations at time-instants 6988 and 14149). However, the steady phase is reached after a further hyperperiod. Again, since no deadline is missed, the task system is schedulable with comprehensive- Earliest Deadline First. It thus follows from Lemma 2, Observation 1 and Observation 2 the conjecture that integer $k \in N^+$ in Expression $(O_{max} + k*P)$ for the time-instant to reach the steady phase must be a function of tasks parameters.

## 5 Conclusions

In this paper, based on this I am presenting a flexible and unified framework for representation of a large family real time tasks and scheduler. Both static and dynamic real time system are covered by this framework. Based on this feasibility interval and considering the predictability property of this scheduler, my main contribution is therefore a precise schedulability test for the comprehensive-Earliest deadline first scheduling of periodic hard real-time tasks upon same multiprocessor platforms. In this paper, we derive a feasibility interval for a Fixed-Job-Priority scheduler, namely comprehensive-Earliest deadline first. In this paper, I am considered the scheduling problem of hard real-time systems composed of periodic constrained deadline tasks upon similar multiprocessor platforms. I am also assumed that tasks were scheduled by using the comprehensive- Earliest Deadline First scheduler and we provided a precise schedulability test for this scheduler. Also, I have showed by means of a counterexample that the feasibility interval, and thus the schedulability test, proposed by Leung [12] is incorrect and also showed which arguments are actually incorrect.

## References

[1] S. Baruah. Techniques for multiprocessor comprehensive schedulability analysis. In Real-Time Systems Symposium, pages 119–128, 20014.

[2] S. Baruah and N. Fisher. The partitioned multiprocessor scheduling of deadline-constrained sporadic task systems. IEEE Trans. on Computers 55 (14), pages 918–923, 2006.

[3] M. Bertogna. Real-Time Scheduling Analysis for Multiprocessor Platforms. PhD thesis, Scuola Superiore Sant'Anna, Pisa, 2008.

[4] M. Bertogna, M. Cirinei, and G. Lipari. Schedulability analysis of comprehensive scheduling algorithms on multiprocessor platforms. IEEE Trans. Parallel Distrib. Syst., 20(4):553–566, 2009.

[5] C. Braun and L. Cucu. Negative results on idle intervals and pe- riodicity for multiprocessor scheduling under Earliest Deadline First. In Junior Researcher Work- shop on Real-Time Computing, 20014.

[6] L. Cucu and J. Goossens. Feasibility intervals for fixed-priority real-time scheduling on uniform multiprocessors. In the 11$^{th}$ IEEE International Conference on Emerging Technologies and Factory Automation, pages 3914–405, 2006.

[8] S. K. Dhall and C. L. Liu. On a real-time scheduling problem. Operations Research, 26:1214–140, 19148.

[9] J. Goossens, S. Funk, and S. Baruah. Priority-driven scheduling of periodic task systems on multiprocessors. Real-time Systems Journal, 25:1814–205, 2003.

[10] R. Ha and J. Liu. Validating timing constraints in multiprocessor and distributed real-time systems. In 14th IEEE International Conference on Distributed Computing Systems, pages 162–1141, 1994.

[11] B. Kalyanasundaram, K. R. Pruhs, and E. Torng. Errata: A new algorithm for scheduling periodic, real-time tasks. Algorithmica, 28:269–2140, 2000.

[12] J.Y.-T.Leung. A new algorithm for scheduling periodic real-time tasks. Algorithmica,4:209–219, 1989.

[13] C. L. Liu and J. W. Layland. Scheduling algorithms for multiprogramming in a hard-real-time environment. Journal of the ACM, 20(1):46–61, 19143.

[14] J.Goossens and P.M.Yomsi Exact Schedubilty Test for Global-Earliest Deadline First Scheduling of periodic Hard Real time System Task on Identical Multiprocessor.

[16] L. Cucu and J. Goossens. Feasibility intervals for multiprocessor fixed-priority scheduling of arbitrary deadline periodic systems. In Design Automation and Test in Europe, pages 1635–1640, 20014.